\documentclass[useAMS,usenatbib]{mn2e}
\usepackage{amsmath}
\usepackage{graphicx}

\usepackage{ enumitem, tabto, xfrac}
\usepackage[]{units}
\usepackage{tabu}
\usepackage{morefloats}
\usepackage{float}
\usepackage{amssymb}
\usepackage{xcolor}
\usepackage[normalem]{ulem}

\title[Low-band Spectral Index with EDGES]{Spectral Index of the Diffuse Radio Background Between 50 and 100~MHz} 

\author[T. J. Mozdzen et al.]{T. J. Mozdzen,$^{1}$\thanks{E-mail:
tmozdzen@asu.edu (TJM)} 
N. Mahesh,$^{1}$
R. A. Monsalve,$^{2,1,3}$
A. E. E. Rogers$^{4}$ and 
J. D. Bowman$^{1}$\\
$^{1}$School of Earth and Space Exploration, Arizona State University (ASU), Tempe, AZ, 85287, USA\\
$^{2}$Department of Physics and McGill Space Institute, McGill University, Montr\'eal, Quebec, H3A 2T8, Canada\\
$^{3}$Facultad de Ingenier\'ia, Universidad Cat\'olica de la Sant\'isima Concepci\'on, Alonso de Ribera 2850, Concepci\'on, Chile\\
$^{4}$MIT Haystack Observatory, Massachusetts Institute of Technology (MIT), Westford, MA, 01886, USA}

\pubyear{2018}

\begin{document}
\label{firstpage}
\date{Accepted 2018 xxx yy. Received 2018 aaa bb; in original form 2018 mmm dd}
\pagerange{\pageref{firstpage}--\pageref{lastpage}} \pubyear{2018}
\maketitle
\begin{abstract}
We report the spectral index of diffuse radio emission between 50 and 100 MHz from data collected with two implementations of the Experiment to Detect the Global EoR Signature (EDGES) low-band system. EDGES employs a wide beam zenith-pointing dipole antenna centred on a declination of $-26.7^\circ$. We measure the sky brightness temperature as a function of frequency averaged over the EDGES beam from 244 nights of data acquired between 14 September 2016 to 27 August 2017. We derive the spectral index, $\beta$, as a function of local sidereal time (LST) using night-time data and a two-parameter fitting equation.  We find $-2.59<\beta<-2.54~\pm$\textcolor{black}{0.011} between 0 and 12~h LST, ignoring ionospheric effects. When the Galactic Centre is in the sky, the spectral index flattens, reaching $\beta = -2.46~\pm$\textcolor{black}{0.011} at 18.2~h. The measurements are stable throughout the observations with night-to-night reproducibility of $\sigma_{\beta}<0.004$ except for the LST range of 7 to 12~h. We compare our measurements with predictions from various global sky models and find that the closest match is with the spectral index derived from the Guzm{\'a}n and Haslam sky maps, similar to the results found with the EDGES high-band instrument for 90-190 MHz. Three-parameter fitting was also evaluated with the result that the spectral index becomes more negative by $\sim$0.02 and has a maximum total uncertainty of 0.016. We also find that the third parameter, the spectral index curvature, $\gamma$, is constrained to $-0.11<\gamma<-0.04$. Correcting for expected levels of night-time ionospheric absorption causes~$\beta$ to become more negative by $0.008$ to $0.016$ depending on LST.
\end{abstract}




\section{INTRODUCTION}

Rapid progress is being made toward observations of the redshifted 21~cm signal from cosmic dawn and the epoch of reionization (EoR).  Our team recently reported evidence for the detection of a 21~cm absorption profile at $z\approx17$ \citep{bow05} in the global sky-average spectrum using the Experiment to Detect the Global EoR Signature (EDGES).  EDGES and the Shaped Antenna measurement of the background RAdio Spectrum 2 (SARAS 2) have both constrained astrophysical parameters related to galaxy evolution during the EoR \citep{mon03,sin01,sin02,sin03} and EDGES has placed a lower limit on the duration of reionziation of $\Delta z\gtrsim 1$ at $z=8.5$ \citep{mon02}.   The Large-Aperture Experiment to Detect the Dark Ages (LEDA) placed an early limit on the global 21~cm spectrum during cosmic dawn \citep{ber02,pri01} and the Dark Ages Radio Explorer (DARE) team has developed new tools for fitting global 21~cm measurements \citep{har01,bur01,tau01,tau02}.  Meanwhile, \cite{phi01} have demonstrated operations of their global 21~cm instrument, Probing Radio Intensity at high-Z from Marion (PRIZM),  in an extremely radio-quiet site. \cite{pat01} have used the High Band Antenna ($7.9<z<10.6$) of the Low-Frequency Array (LOFAR) to constrain the power spectrum of 21~cm spatial fluctuations to within an order of magnitude of the expected signal during reionization and, \cite{geh01} have begun to constrain the 21~cm power spectrum during cosmic dawn using the Low Band Antenna of LOFAR.  The Precision Array for Probing the Epoch of Reionization (PAPER) team are preparing their final constraints on the 21~cm power spectrum after correcting and improving their pipeline \citep{ali01,ali02,che01}. The Hydrogen Epoch of Reionization Array (HERA) is coming online \citep{deb01}, and the Murchison Widefield Array (MWA) plans to build on its initial power spectrum constraints \citep{bea01,ewa01} with its recent Phase II extension \citep{way02}.  

Central to all of these efforts is foreground mitigation and subtraction. Advances in the accuracy of sky models and improved foreground removal techniques are necessary to separate the 21~cm signal from the Galactic and extragalactic foregrounds, which are up to five orders of magnitude stronger.   HERA, LOFAR, MWA, and others continue to address foreground removal issues such as beam chromaticity, foreground modelling strategies, foreground avoidance, foreground source removal, and widefield considerations \citep{car01,cha01,chl01,mck01,pob01,pro01, thy01,thy02,ker01}  For EDGES, an accurate diffuse foreground model is needed to correct chromaticity in the antenna's beam response on the sky.  Yet much of our knowledge about the properties of the radio sky below 200~MHz still relies heavily on the early pioneering measurements from up to five decades ago \citep{alv01,bri01,has02,mae01,tur01}, or is simply extrapolated from other frequencies as is often done using the 408~MHz all-sky map of \cite{has02}.

Several efforts have recently been undertaken to improve our knowledge of the radio sky below 200~MHz by expanding and improving maps of diffuse emission and catalogs of radio sources.  A new analysis by \cite{zhe01} improves on the Global Sky Model (GSM) of \cite{deo01} by including data from 29 measured sky maps instead of the original 11, including two maps acquired with the Parkes telescope \citep{lan01} at 85 MHz and 150 MHz. The new GMOSS sky model \citep{rao01,rao02} uses four measured sky maps between the frequencies of 150 MHz and 23 GHz along with two GSM generated maps at 22 and 45 MHz and applies the physics of radiative processes to aid accuracy and consistency. The  45~MHz surveys of \citet{alv01} and \citet{mae01} have been revisited in the \citealt{guz01} all-sky temperature map.  The Long Wavelength Array (LWA)-1 Low Frequency Sky Survey \citep{dow01} and the Owens Valley-LWA sky model \citep{eas01} provide new references for the northern hemisphere sky.  \cite{deg01} have created a spectral index sky map covering 80\% of the sky using sky surveys at 147 and 1400~ MHz.  A recent workshop on the Radio Synchrotron Background \citep{sing01} highlighted the need for a new all-sky diffuse radio background survey to refine our present maps. Source catalogs are also improving with recent contributions from the LOFAR Two Metre Sky Survey \citep{hea01} and the LOFAR Multifrequency Snapshot Sky Survey \citep{shi01} in the northern latitudes, the GaLactic and Extragalactic All-sky MWA (Murchison Widefield Array) (GLEAM) \citep{way01} in the lower latitudes, and the Giant Metrewave Radio Telescope (GMRT) \citep{int01}.

The EDGES instrument can contribute to the knowledge of the diffuse radio sky by analyzing the sky spectra it collects while searching for the global 21~cm signal. The wide beam of the EDGES instrument averages the absolute sky brightness temperature over large spatial scales on the sky. Past measurements using EDGES determined the spectral index at high Galactic latitudes to be $\beta_{150-408} = -2.52\pm0.04$  \citep{rog01}.  In \cite{moz01} we obtained a refined measurement of the spectral index in the frequency range 90-190~MHz over the full local sidereal time (LST) range. We found $-2.60>\beta>-2.62~\pm$0.02 between 0 and 12~h LST, and when the Galactic Centre is in the sky, the spectral index flattened to $\beta = -2.50~\pm$0.02 at 17.7~h. The measurements provide a well-calibrated anchor against which sky maps (particularly from interferometers that are not sensitive to total power) can be tied to absolute flux scales and spectral shape. Based on the work of \cite{pla01} who showed that in the  $1-10$~GHz frequency range, the spectral index flattens towards lower frequencies and more recently \cite{deo01} reported a similar trend at 150~MHz and 5~GHz, we expect the spectral index as measured by EDGES to flatten as the frequency range shifts from $90-190$~MHz to $50-100$~MHz.

In this paper we present new observations with two low-band versions of the EDGES instrument operating between 50 and 100~MHz.  We compare our low-band measurements of the spectral index with the results found for the measured data using the high-band antenna \citep{moz01} and with the computed spectral index using the low-band antenna beam along with the GSM, the improved GSM, the GMOSS, and the Guzm{\'a}n/Haslam sky models. We evaluate the spectral index using both two and three-parameter fitting, and we calculate the impact of ionospheric effects.  The paper is organized in the following manner.  In Section~2 we briefly describe the instrument, and in Section~3 we present details of the data collected, calibrations, beam chromaticity correction, as well as the modelling of the sky spectrum. Section~4 presents and discusses the spectral index results, including comparisons to values predicted by relevant sky models. Section 5 contains the conclusions of this work.
\section{EDGES INSTRUMENT}
\label{sec:instrument}

The EDGES experiment is deployed at the Murchison Radio-astronomy Observatory (MRO) in Western Australia (-26.7$^\circ$,~+116.6$^\circ$).  Several versions of the EDGES instruments have been operated since 2015.  Two low-band instruments observed frequencies from 50 to 100~MHz and one high-band instrument observed from 90 to 190~MHz. Each instrument consists of a single dipole-based antenna and a temperature controlled calibrated receiver.  The details of the low-band instruments, their calibration, and the methods to estimate total uncertainty are very similar to the high-band instrument that was discussed in detail in \cite{rog02, mon01, moz01}.    The primary difference between the low-band instruments and the high-band is the size of the antenna and ground plane for each instrument.

\subsection{Antenna and ground plane} 

\begin{figure}
  \includegraphics{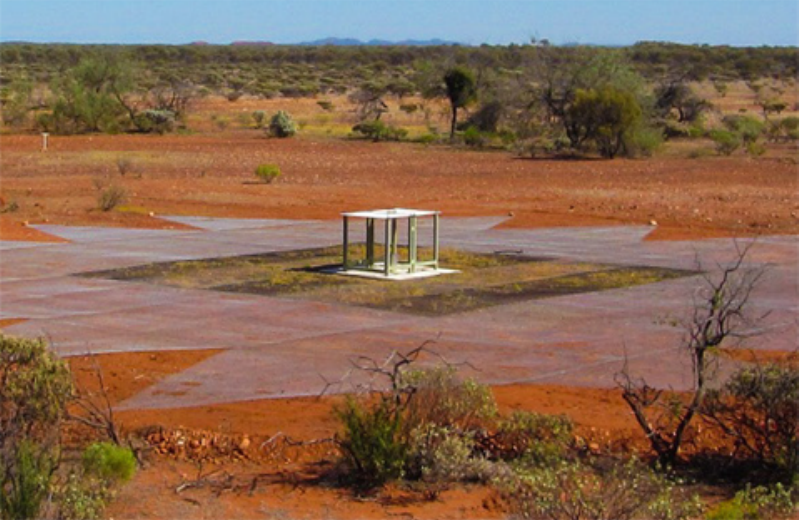}
 \caption{Photograph of the Lowband 1 antenna and its ground plane.  The horizontal ``blade'' dipole antenna is supported by fiberglass legs.  Two brass balun tubes connect the antenna to the ground plane.  One of the balun tubes provides a coaxial transmission line for propagating signals to the receiver, which is installed under the ground plane.  A small metal balun shield is installed at the base of the balun tubes to reduce horizon response.  The ground plane is formed from panels of metal wire mesh welded together (except for the inner $2.2\times2.2$~m$^2$, which is formed from solid metal panels) and extends to 30~m edge-to-edge.}
  \label{fig:antenna_images}
\end{figure}

\begin{figure}
  \includegraphics{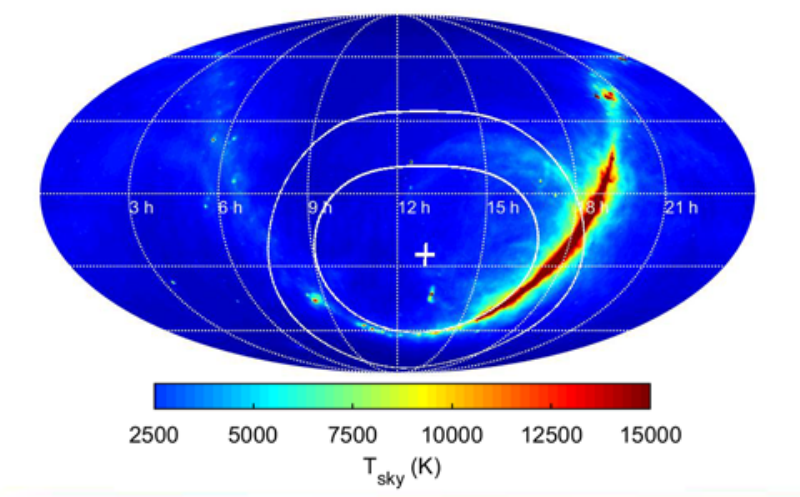}
 \caption{The EDGES beam at -3 and -10 dB projected onto the Haslam sky map scaled to 75~MHz in Celestial coordinates.  Due to the latitude at the MRO, the zenith-centred beam is centred at -26.7$^\circ$ declination. In this example the map is projected with 13~h LST at zenith. The beam is purposefully wide to capture the sky-averaged signal.}
  \label{fig:beam_width}
\end{figure}

The low-band instruments employ a broadband ``blade'' dipole antenna sensitive to wavelengths in the range $6.0~\geq~\lambda~\geq~3.0$~m ($50\le\nu\le~100$~MHz) and are based on scaling the high-band antenna by a factor of two.  Each antenna is ground-based and zenith pointing and consists of two rectangular planar panels that are each 125.2~$\times$~96.4~cm$^2$, and placed 104~cm above a ground plane.  The ground plane consists of a 2.2$\times$2.2~m$^2$  aluminum sheet centered underneath the antenna, whose perimeter attaches to sheets of wire mesh that form a 20$\times$20~m$^2$ square. Additional extensions of mesh triangles, 5~m~base~$\times$~5~m~height, are attached along its perimeter to form the final structure with a total span of 30$\times$30~m$^2$ (see Fig.~\ref{fig:antenna_images}).  A balun consisting of two parallel metal tubes connects the antenna panels to the ground plane and conducts the sky signal to the receiver, which is installed under the ground plane.  A small metal balun shield is located at the base of the tubes to reduce sensitivity of the antenna to the horizon. The full width at half maximum beamwidth of the antenna at 75~MHz is 71.6$^\circ$ parallel to the axis of the dipole excitation and 108$^\circ$ perpendicular to this axis. Fig.~\ref{fig:beam_width} shows -3dB and -10dB limits of the beam's directivity projected onto the Haslam sky map scaled to 75~MHz in Celestial coordinates. The properties of the antenna are summarized in Table~\ref{tab:antennas} and a deployed antenna can be seen in Fig.~\ref{fig:antenna_images}.

\begin{table}
   \caption{Blade Antenna Properties (refer to Fig. \ref{fig:antenna_images})}
   \label{tab:antennas}
   \begin{tabular} {l  l}
   	\hline
  	Parameter & Value\\
   	\hline
  	3 dB Beamwidth, $\phi$=~0$^{\circ}$ at 75 MHz &71.6$^{\circ}$\\
  	3 dB Beamwidth, $\phi$=90$^{\circ}$ at 75 MHz &108$^{\circ}$\\
 	Height above ground plane &104 cm\\
	Panel Width &125.2 cm\\
	Panel Length &96.4 cm\\
	Solid ground plane&2.2~$\times$~2.2 m$^2$\\
	Inner square mesh extension &20~$\times$~20 m$^2$\\
      Outer triangular mesh extensions &$5\times5$~m$^2$ each\\
     \hline
	\end{tabular}
\end{table}

\section{DATA AND PROCESSING}
\subsection{Observations}
We measured the sky temperature using EDGES for 244 nights over a period of 348 days starting with day 258 (September 14) of 2016 and ending with day 239 (August 27) of 2017.  The first data set was collected by the instrument labeled Lowband 1, whose antenna was oriented in a North-South (NS) direction, and ran for 126 nights between day 258 2016 through day 17 2017. The second data set was collected by Lowband 2.  For 61 nights, Lowband 2 observed using a NS oriented antenna.  Its antenna was then rotated to an East-West (EW) orientation, and data were collected for 17 more nights. Lastly, its balun shield was removed, and data were collected for an additional 58 nights, still in an EW orientation. There were periods when the instruments were unavailable due to either weather issues or equipment down-time.  Table~\ref{tab:summarized_data} summarizes the observing program.

Only night-time (Sun elevation $<-10^{\circ}$) data are included to minimize solar and ionospheric disturbances to the astronomical sky foreground \citep{rog03,ved01}.  Some frequency channels in observed spectra are contaminated by radio-frequency interference (RFI). These channels are flagged and excised. Obvious anomalies found after the fitting process, described in Section \ref{sec:modelling}, are also excised. The data are averaged and binned each night into 72~time slots of 20~minutes in LST and into 125 frequency bins of 400~kHz from 50~to 100~MHz. \textcolor{black}{These 20 minute observations consist of approximately 30 individual 39~s measurement cycles divided evenly between the sky and two internal references. Considering that the digitizing efficiency is ~18\%, only 72~s of effective sky integration time is produced per 20 minute bin.}

We calibrate the data using the technique presented in \cite{rog02} and \cite{mon01}, which includes field and lab characterization measurements. During field calibration the sky measurement is combined with data from hot and cold internal reference loads which increases the thermal noise in the sky measurements by a factor of $\sim$2. The absolute calibration parameters used for the low-band instruments are shown in the Methods section of \cite{bow05}.

\subsection{Adjustment for beam chromaticity} 
Similarly to \cite{moz01}, we correct the measured spectra for chromaticity using electromagnetic simulations of the beam. The blade dipole was modelled using FEKO's Method of Moments solver. To capture the effects of soil below the ground plane, we included a dielectric medium in the lower half of the simulation space (below ground) with a relative permittivity of $\epsilon_r = 3.5$ and conductivity of $\sigma~=~2~\times 10^{-2}~S/m$, which is based on soil measurements made at the MRO for dry conditions \citep{sut01}. The simulation frequency limits were 50 and 100~MHz with a 1~MHz resolution, and the angular resolution of the beam's directivity was 1~degree.

\begin{figure}
 \centering
 \includegraphics[trim=0 22 0 0, clip,width=3.3 in]{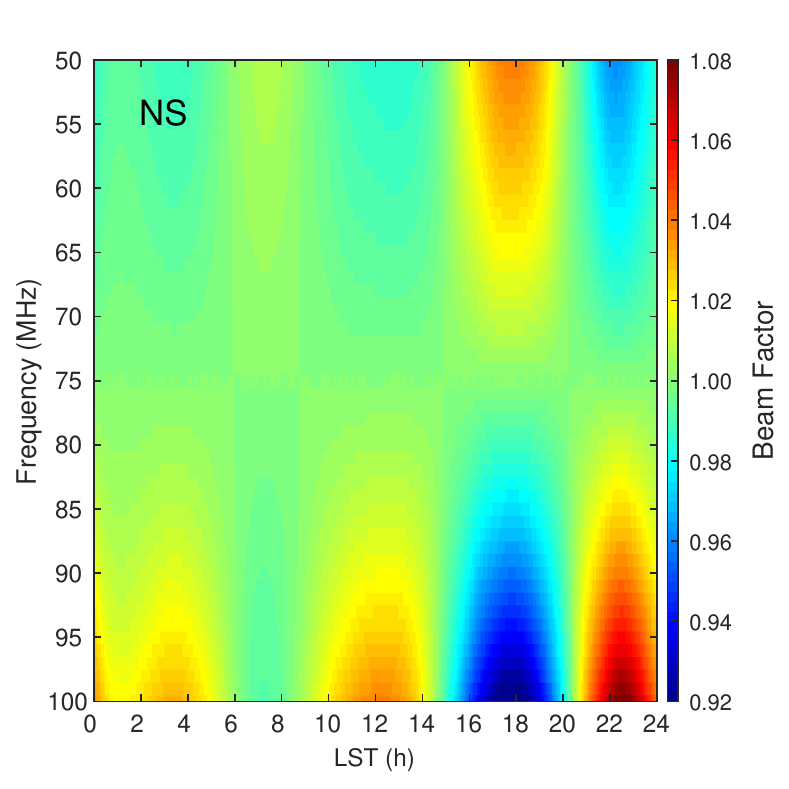}
 \includegraphics[trim=0 0 0 0, clip,width=3.3 in]{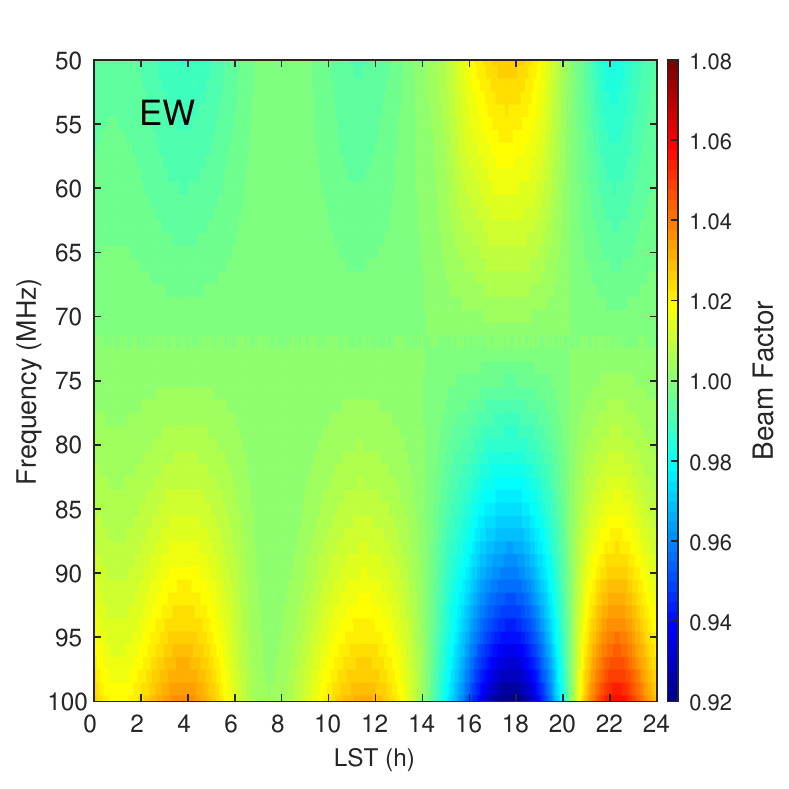}
 \caption{\textcolor{black}{The beam chromaticity correction factors are shown for (top) an NS dipole excitation axis and (bottom) an EW excitation axis. The factors were computed by scaling the Haslam 408 MHz sky map to 75~MHz, simulating the antenna temperature at each LST using the antenna's beam at 1~MHz intervals between 50 and 100~MHz, and calculating the ratio of antenna temperature at each frequency point relative to the antenna temperature at 75~MHz. Measured spectra are modified to compensate for this expected chromatic variation in antenna temperature by dividing by the correction factor.}}
 \label{fig:BeamFactors}
\end{figure}

For a given LST value, the beam correction factor is obtained using the simulated beam solutions and the 408~MHz Haslam all-sky map \citep{has02} with the following equations: 
\begin{equation} 
\label{eq:beam_correction}
B_{\text{factor}}(\nu) = \frac{\int_{\Omega}T_\text{sky-model}(\nu_{_{75}},\Omega) B(\nu,\Omega) \mathrm{d}\Omega}{\int_{\Omega}T_\text{sky-model}(\nu_{_{75}},\Omega) B(\nu_{_{75}},\Omega) \mathrm{d}\Omega },
\end{equation}
where
\begin{equation} 
\label{eq:sky_correction}
T_\text{sky-model}(\nu_{_{75}},\Omega) = \left [ T_\text{Haslam}(\Omega) - T_\text{CMB}\right ] \left ( \frac{75}{408} \right )^{-2.5}  + T_\text{CMB},
\end{equation}
$B(\nu_{_{75}}, \Omega)$ is the beam directivity at 75~MHz for a given pointing and orientation, $\nu$ is frequency, $\Omega$ are the spatial coordinates above the horizon, $T_\text{Haslam}(\Omega)$ is the 408~MHz Haslam all-sky map, and the CMB temperature, $T_\text{CMB}$, is 2.725~K. Unique beam correction factors are generated for each of the 72~LST time intervals. The measured sky spectra are then corrected by dividing by the beam correction factor. \textcolor{black}{Separate factors are computed for NS and EW excitation axis orientations and are shown in Fig.~\ref{fig:BeamFactors}.}

\textcolor{black}{Deviations from the ideal assumptions that the spatial structure of the foreground is the same at 75~MHz as it is at 408~MHz and that the antenna's beam directivity is known exactly leads to additional uncertainty when calculating the spectral index. These effects are discussed later in section \ref{sec:BeamFactor}.}

\subsection{Modelling}
\label{sec:modelling}
Since the sky foreground is dominated by Galactic synchrotron radiation, we expect to be able to model its temperature as a power law series  with a base exponent near $-2.5$. Modeling the foreground in this manner has been shown to be adequate by, e.g., \cite{deo01,rog01,bow02,voy01,patr01,moz01a,ber02,bow05, dow02, mck01}.

We choose to focus on \textcolor{black}{a two fitting term and a three fitting term expansion} to determine the spectral index by fitting the calibrated data, $T_\text{ant}$, to the two-parameter equation
\begin{equation} 
\label{eq:Power_Law_2P}
T_\text{ant} = T_{75}\left(\frac{\nu}{\nu_{_{75}}}\right)^{\beta} + T_\text{CMB},
\end{equation} 
and the three-parameter equation
\begin{equation} 
\label{eq:Power_Law_3P}
T_\text{ant} = T_{75}\left(\frac{\nu}{\nu_{_{75}}}\right)^{\beta+\gamma \ln(\frac{\nu}{\nu_{_{75}}})} + T_\text{CMB},
\end{equation} 
where the fitting parameters, $T_{75}$, $\beta$, and $\gamma$ are the brightness temperature at 75~MHz less $T_\text{CMB}$, the spectral index, and the second order spectral index curvature parameter respectively. The fitting algorithm used was Levenberg-Marquardt as implemented by Mathematica's NonLinearModelFit function. For our primary analysis we neglect ionospheric effects. We will examine the effects of the ionosphere in section~\ref{sec:Ionosphere}. 

\begin{table}
  \center
   \caption{Data were collected for 244 unique nights over the span of nearly one year between day 258 of 2016 and day 239 of 2017. The periods of data collection from both low-band instruments are indicated in the table.}
   \label{tab:summarized_data}
   \begin{tabular} {l  l  r c}
   	\hline
  	Instrument configuration&Year&Day Numbers&Span\\
   	\hline
	 Lowband 1 NS&2016&258~to~366  &109\\
	Lowband 1 NS&2017&~~001~to~017& ~17\\
    Lowband 2 NS&2017&~~082~to~142 &~61\\
    Lowband 2 EW&2017&~~155~to~171 &~17\\
    Lowband 2 EW, no shield&2017&~~181~to~239 &~58\\
     \hline
	\end{tabular}
\end{table}

\section{RESULTS}
\label{sec:results}
\subsection{Spectral index} 
Spectral data were fit separately per day for each LST interval (see Table ~\ref{tab:summarized_data}). Both two-parameter fitting (equation~(\ref{eq:Power_Law_2P})) and three-parameter fitting (equation~\ref{eq:Power_Law_3P}) were applied to the calibrated and beam-corrected spectra. 

\subsubsection{Two-parameter fitting}
\begin{figure*}
 \begin{minipage}[b]{0.99\linewidth}
\centering
   \includegraphics[trim=00 0 0 0, clip,width=3.3 in]{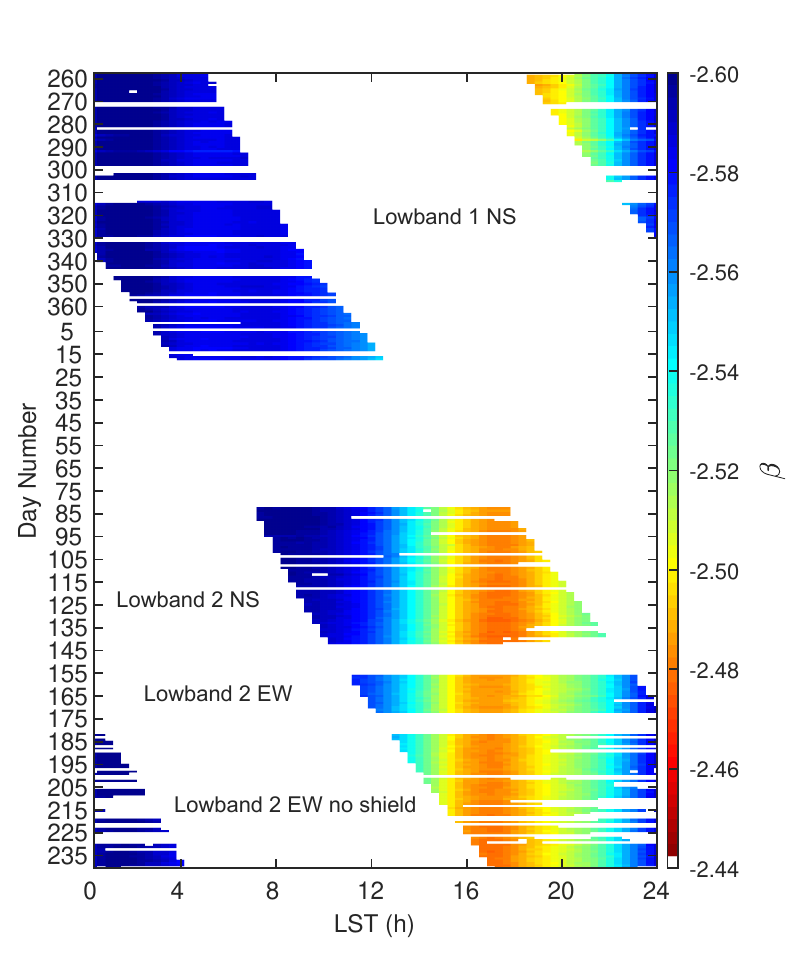}
\hspace{0.5 cm}
   \includegraphics[trim=25 0 0 0,  clip,width=2.94 in]{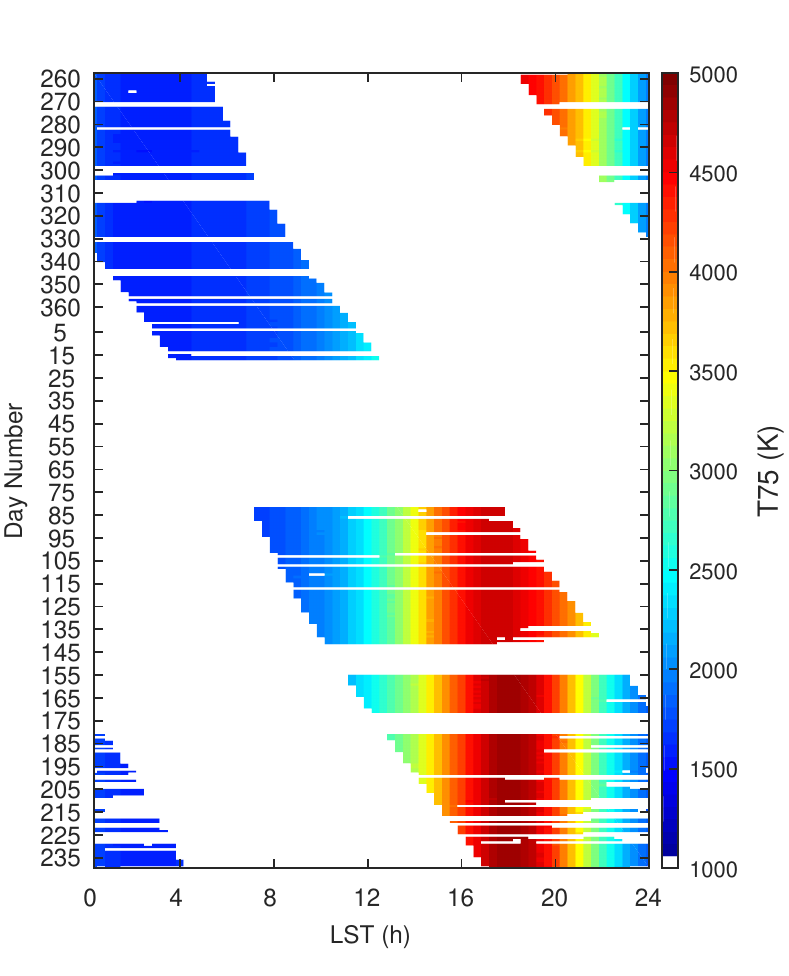}
 \end{minipage}
\begin{minipage}[b]{0.99\linewidth}
\vspace{0.25 cm}
\centering
   \includegraphics[trim=0 0 0 18, clip,width=3.3 in]{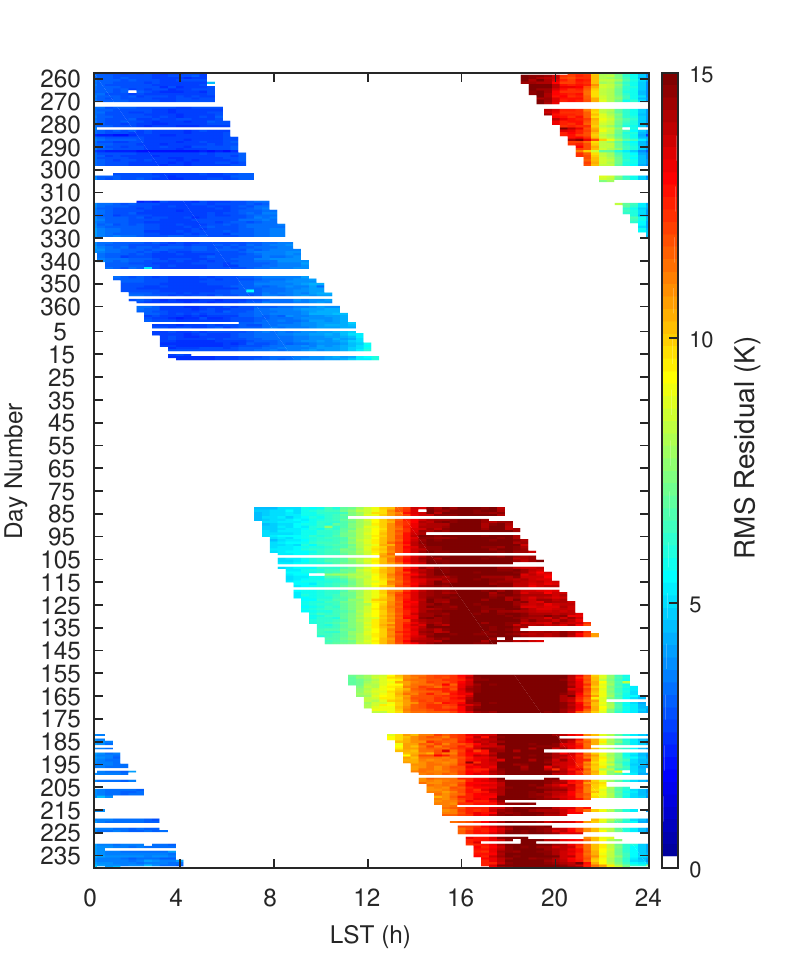}
   \caption{Two-parameter waterfall graphs of the (top-left) spectral index $\beta$, (top-right) sky temperature at $\nu$=75~MHz, $T_{75}$, and (bottom) the RMS residual to equation~(\ref{eq:Power_Law_2P}). RFI and other spurious signals were purged from the spectra before performing the fit. The data were binned into 400~kHz wide bins from 50 to 100 MHz and corrected for beam chromaticity. Daytime data (sun above $-10^\circ$ elevation) were not analysed, yielding the blank diagonal bands. The results show excellent stability from day to day as the data collection ran for a span of 348 days from day 258 2016 to day 239 2017, of which 244 nights were usable.  The EW antenna orientation of the last two data sets yields a slight increase in $T_{75}$ and a shift in the location of maximum RMS residual due to the antenna beam's slightly different weighting of the spatial structure in the sky compared to the NS orientation. Table \ref{tab:summarized_data} lists the dates of operation for the four instrument configurations: Lowband 1 NS covers days 258 2016 through 17 2017; Lowband 2 NS ran from day 82 to 142 2017; Lowband 2 EW ran from day 155 to 171 2017; and Lowband 2 EW without the balun shield ran from day 181 to day 239 2017.}
\label{fig:Waterfalls_two_params}
 \end{minipage}
\end{figure*}

\begin{figure}
\centering
   \includegraphics[trim= 0 21  0 5, clip,width=3.3 in]{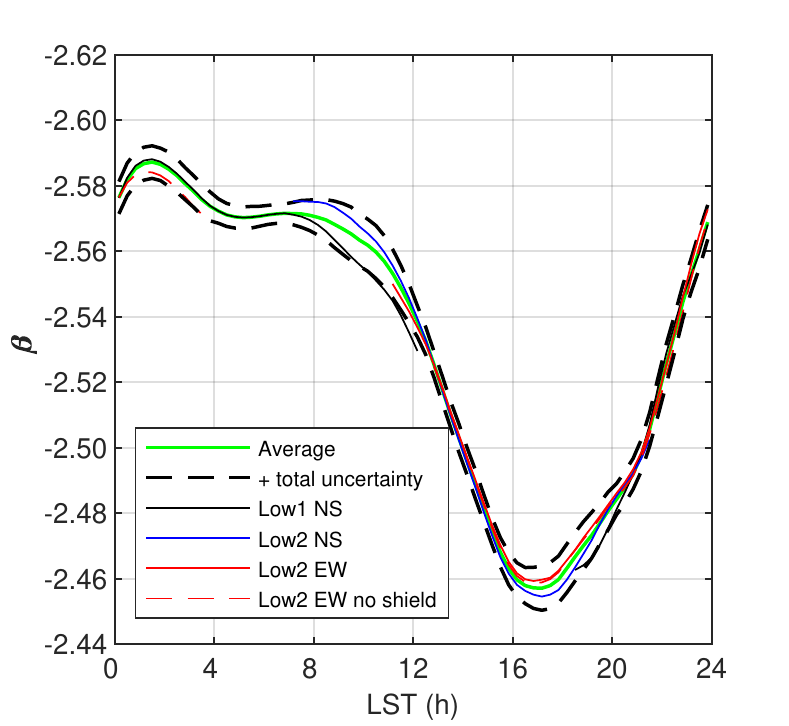}
   \includegraphics[trim= 0 21  0 5, clip,width=3.35 in]{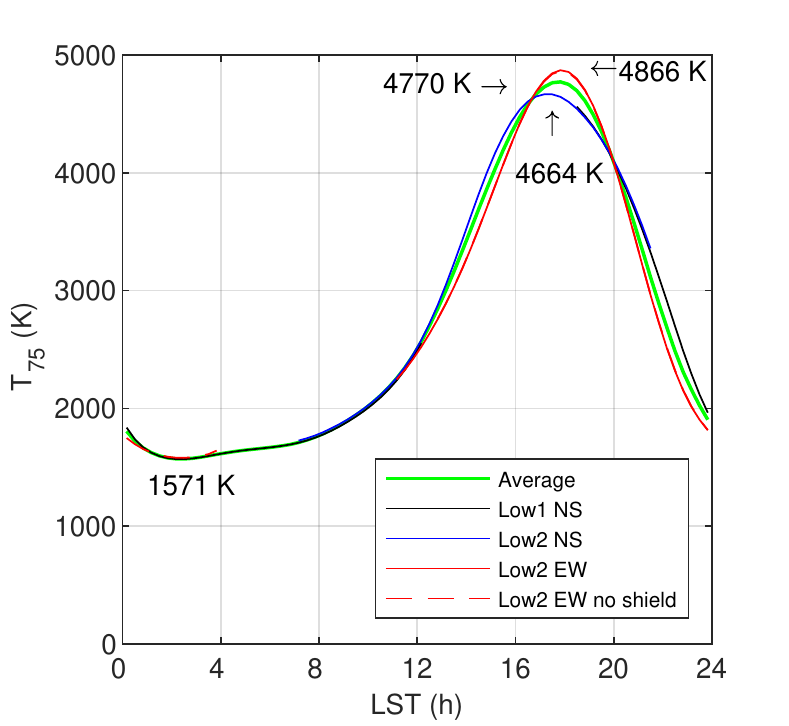}
   \includegraphics[trim= 0 00 10 5, clip,width=3.15 in]{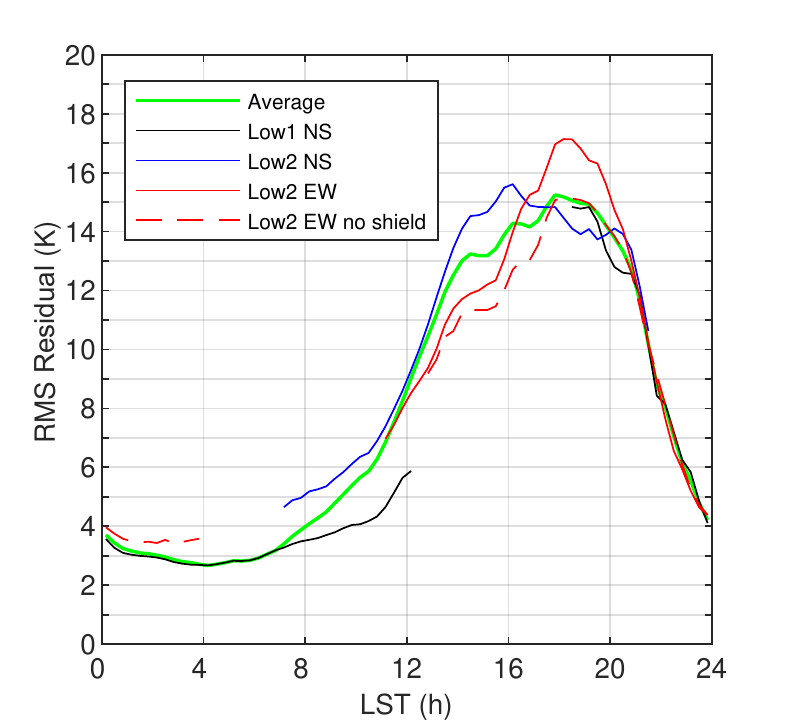}
      \caption{Two-parameter fit averages of $\beta$, $T_{75}$, and the RMS residual for each LST value for each of the instrument configurations taken over days for which night-time data were available. The average of all data sets is shown as the green curve and the 1$\sigma$ total uncertainty for \textcolor{black}{20 minute observations} is shown as a dashed line in the $\beta$ plot. The highest data scatter for the spectral index occurs between LST values of 7 and 12 h.}
\label{fig:Grouped_Averages_two_params}
\end{figure}

The parameters $T_{75}$ and $\beta$ were extracted by fitting the data to equation~(\ref{eq:Power_Law_2P}). The best-fit values of $T_{75}$ and $\beta$, as well as the RMS residuals, are presented in the waterfall plots of Fig.~\ref{fig:Waterfalls_two_params}. They show good stability over time within a given instrument. The averages from the individual data sets for each parameter are plotted in Fig.~\ref{fig:Grouped_Averages_two_params}, along with the combined average between datasets, which we present as a reference, for each parameter across all data sets. For $\beta$ we also plot the total uncertainty for each of the 72 LST bins, which is the sum of the standard deviation of the data scatter, for 20 minute \textcolor{black}{observations}, between the days used for the average and an estimate of the systematic uncertainty, discussed in section~\ref{sec:error} below. In general, the dominant uncertainty between LST values of 7 and 12 h is from instrument to instrument differences and outside this range, from systematic (calibration) uncertainty.   The spectral index results of the two systems agree with each other across LST, differing by less than 0.005 except in the range from 8 to 12 h, where they differ by up to 0.01.  This is also the region with the least amount of data. The peak and minimum values for $T_{75}$ are annotated on the graph for each antenna orientation and the combined average.  \textcolor{black}{The antenna orientation, NS or EW, causes a two bin shift in the location of the $T_{75}$ peak and a magnitude difference of 202~K.  This is expected because the two orientations of the antenna are sensitive to slightly different portions of the sky due to the asymmetric shape of the beam (see Table~\ref{tab:antennas}). These effects were also seen in simulations using the beam directivity and the scaled 408~MHz Haslam sky map, where the $T_{75}$ peak using the NS oriented antenna occured 0.5 h earlier than the EW oriented antenna, and was lower by $\sim$225~K.} There is no noticeable comparable effect for $\beta$ because proportionally, it varies much less with LST than the sky temperature.  In comparison to the spectral index measurements made with the high-band instrument between 90 and 190~MHz \citep{moz01}, the spectral index values found here between 50 and 100~MHz are less negative (shallower) by 0.03~to~0.06 across the LST range. This will be discussed more in section \ref{sec:sim}.

\textcolor{black}{In this work, we assume all observations are adequately described by unpolarized sky models.  EDGES samples only a single polarization in any given configuration and could potentially confuse polarized sky signals with the overall unpolarized intensity.  However, we expect any polarized contribution to be very small at the 5-50~mK level based on polarized surveys such as \cite{len01} and the fact that the large beam of EDGES averages over any strucutre.} 

\subsubsection{Three-parameter fitting}
\begin{figure*}
 \begin{minipage}[b]{0.99\linewidth}
\centering
   \includegraphics[trim=0 28 0 0, clip,width=3.3 in]{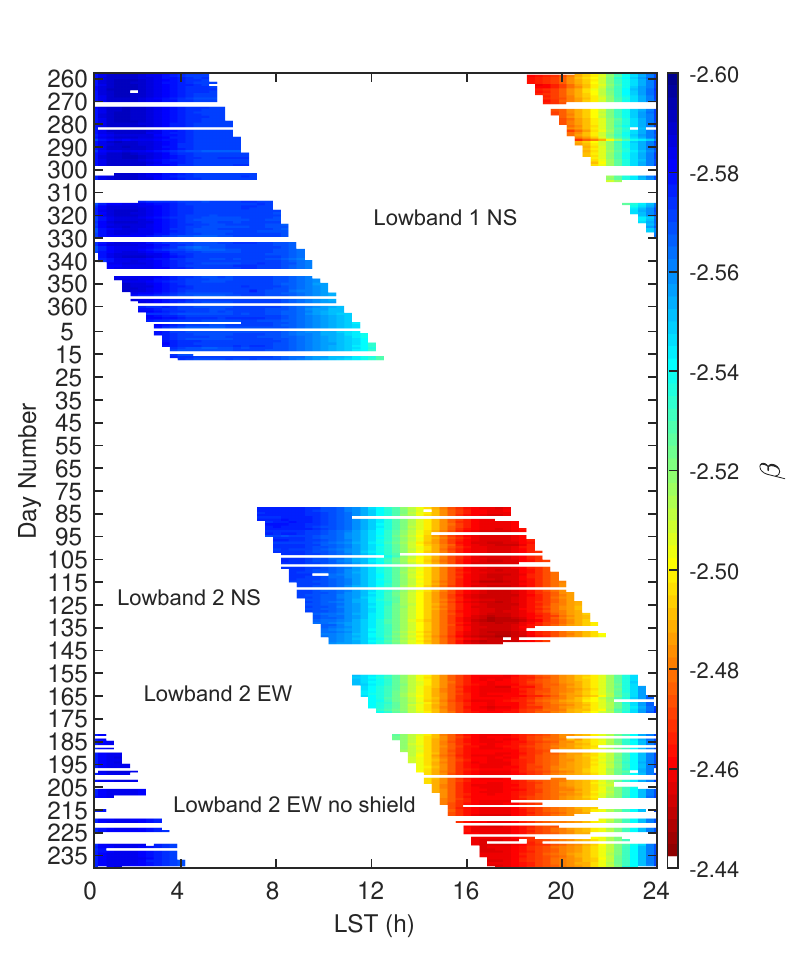}
\hspace{0.5 cm}
   \includegraphics[trim=26 28 0 0, clip,width=2.94 in]{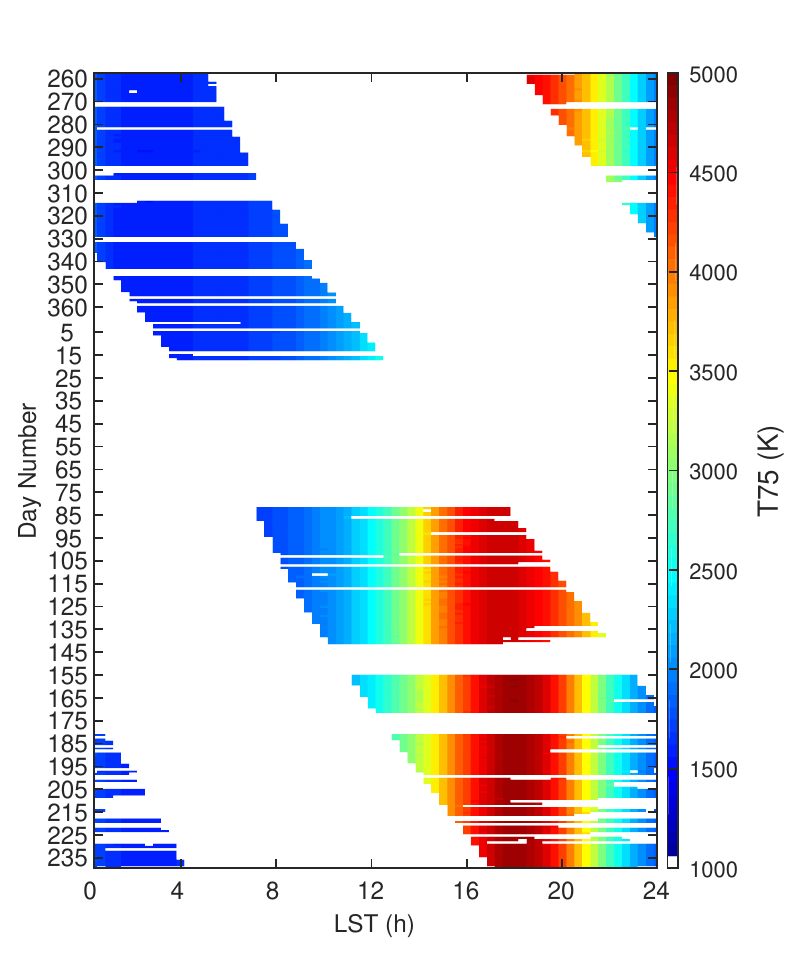}
 \end{minipage}
\begin{minipage}[b]{0.99\linewidth}
\vspace{0.10 cm}
\centering
   \includegraphics[trim=0 0 0 0, clip,width=3.3 in]{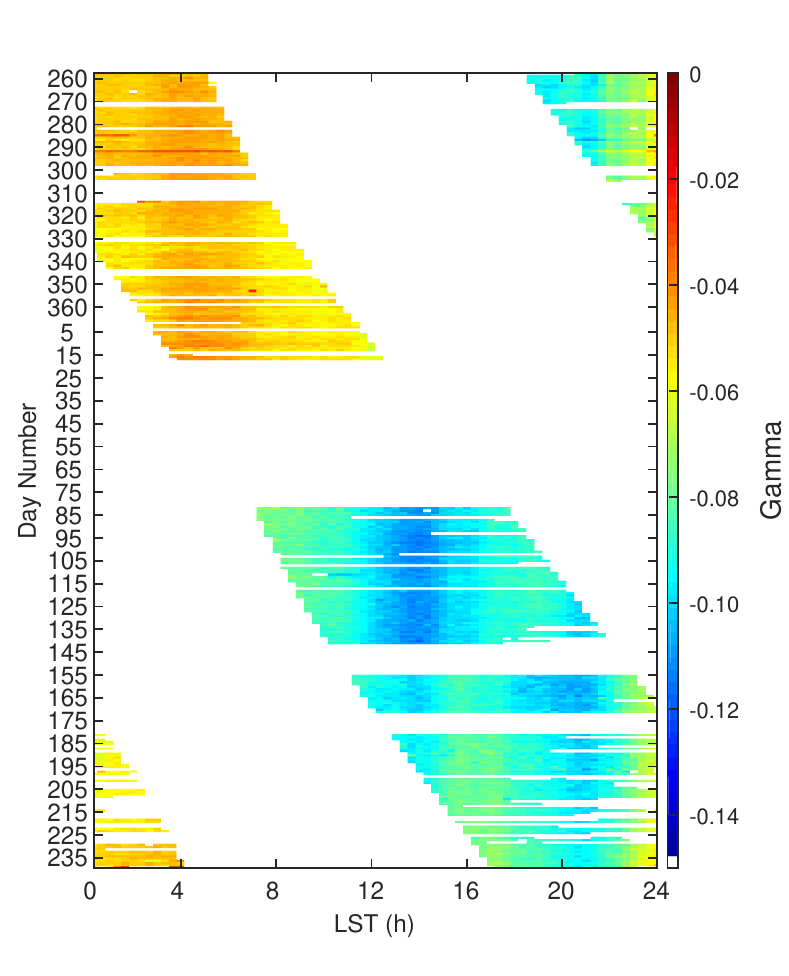}
\hspace{0.5 cm}
   \includegraphics[trim=25.5 0 0 0, clip,width=2.94 in]{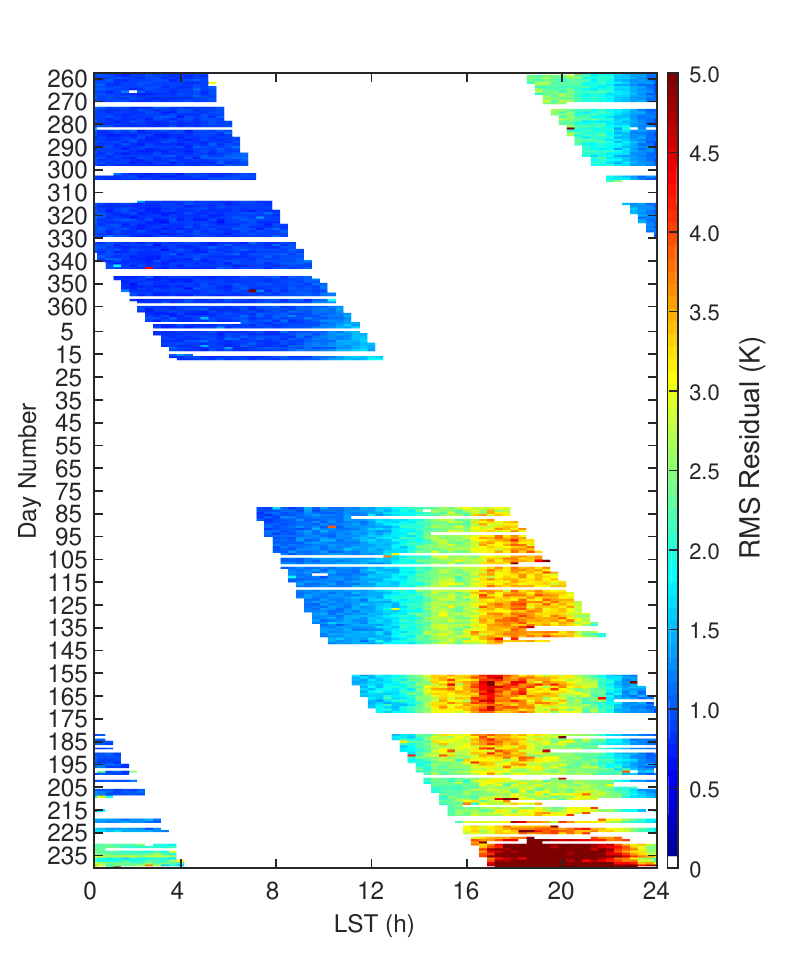}
   \caption{Three-parameter waterfall graphs of the (top-left) spectral index $\beta$, (top-right) sky temperature at $\nu$=75~MHz, $T_{75}$, (bottom-left) spectral index curvature $\gamma$, and (bottom-right) RMS residual to equation~(\ref{eq:Power_Law_3P}). Day to day stability of the instrument is again indicated in the graphs. Compared to the results from two-parameter fitting, the spectral index has steepened by approximately 0.02, the RMS residual levels have decreased in general with the maximum value falling by a factor of $\sim$3, and $T_{75}$ is largely unchanged. The new third fitting parameter, $\gamma$, is seen to lie between $-0.04~\text{and} -0.11$}

\label{fig:Waterfalls_three_params}
 \end{minipage}
\end{figure*}

\begin{figure}
\centering
   \includegraphics[trim= 0 21  0 5,clip,width=3.20 in]{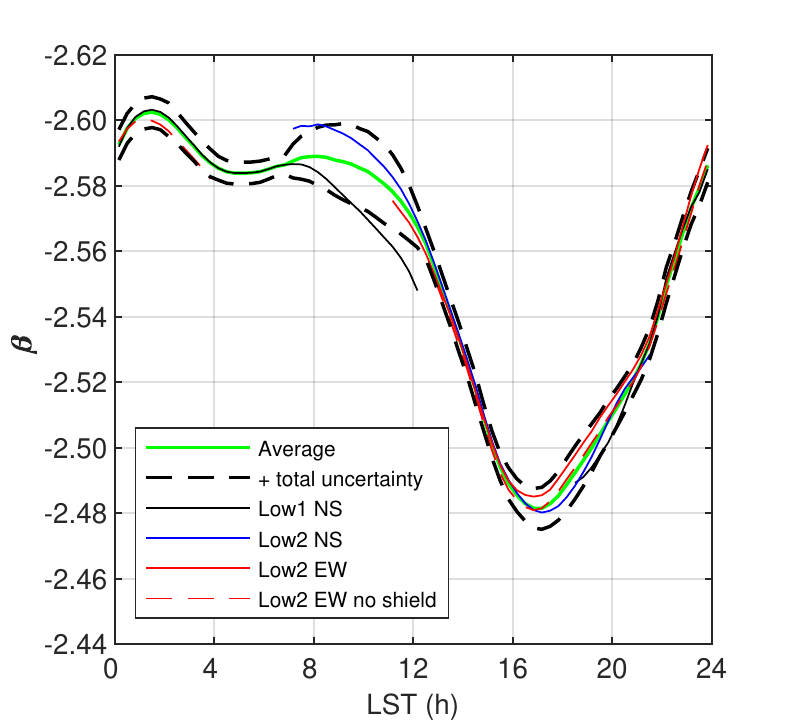}
   \includegraphics[trim= 0 21  0 5,clip,width=3.20 in]{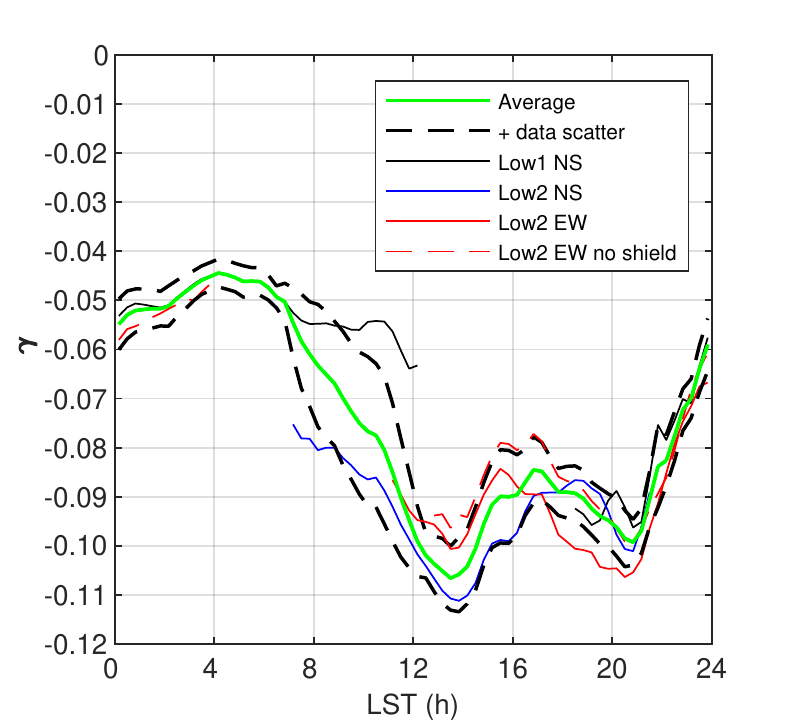}
   \includegraphics[trim= 0 00  5 5,clip,width=3.15 in]{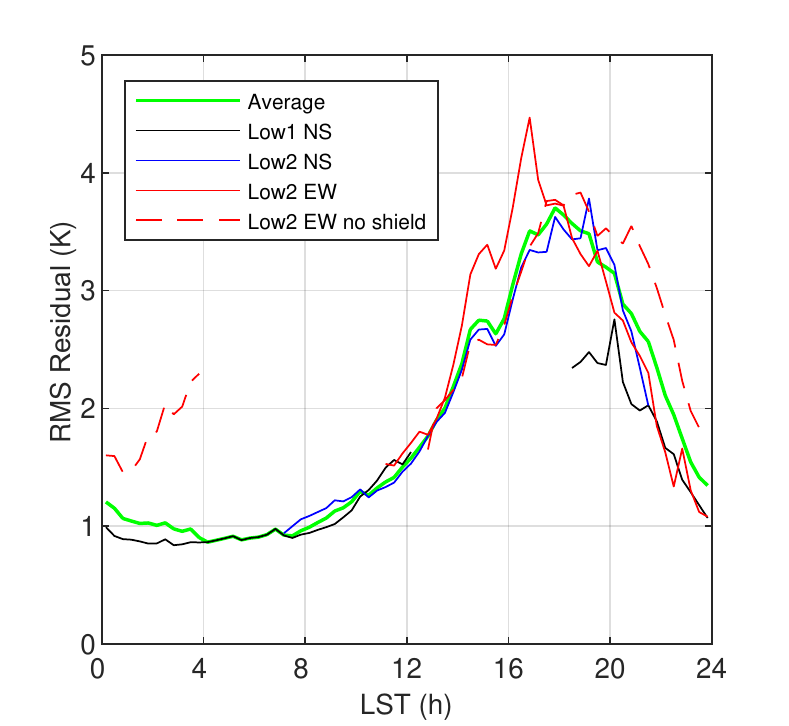}
      \caption{Three-parameter fit averages of $\beta$, $\gamma$, and the RMS residual for each LST value for each of the instrument configurations taken over days for which night-time data were available. The average of all data is shown as the green curve and the 1$\sigma$ total uncertainty for \textcolor{black}{20 minute observations} is shown as a dashed line in the $\beta$ plot. The highest data scatter for the spectral index occurs between LST values of 7 and 12 h. The values for $T_{75}$ differ from the two-parameter values by no more than 4~K and are not shown for brevity.}
\label{fig:Grouped_Averages_three_params}
\end{figure}

For three-parameter fitting, the parameters $T_{75}$, $\beta$, and $\gamma$ were extracted by fitting the data to equation~(\ref{eq:Power_Law_3P}). Again, the best-fit values of $T_{75}$, $\beta$, and $\gamma$, as well as the RMS residuals, show good stability over time within a given instrument, but they are more dependent on the particular instrument configuration as can be seen in the waterfall plots of Fig.~\ref{fig:Waterfalls_three_params}.   The averages from the individual data sets for each parameter and the combined average for each parameter across all data sets are plotted in Fig.~\ref{fig:Grouped_Averages_three_params}.  The plot for $T_{75}$ is not shown because it is virtually identical to the results from two-parameter fitting. The impact of adding a curvature parameter to the fit model is that the spectral index becomes more negative by approximately 0.02 as compared to two-parameter fitting.   The curvature parameter ranges between $-0.11~<~\gamma~<~-0.04$, with the more negative values occurring near the Galactic Centre. These values are more positive than the values estimated in \cite{deo01} at 150~MHz, where $\gamma$ ranges mostly from $-0.30$ near the Galactic Centre to $-0.12$ at high galactic latitudes.

\subsection{Systematic uncertainty estimation}
\label{sec:error}
In the two- and three-parameter fits discussed above, we estimated the impact of errors in the calibration from the following effects: ground loss, antenna loss, and beam chromaticity. As with the high-band instrument, we maintain the receiver at $25^{\circ}$C with forced airflow and active cooling/heating. No temperature offset correction between the main temperature sensor and the LNA was used in this analysis because the departures from the control temperature are typically $< 1^{\circ}$C throughout the season and the impact on the estimated parameters is negligible compared to the other effects discussed.

\subsubsection{Ground loss}
Given the limited size of the metal ground plane ($\sim$~7~wavelengths at 75~MHz), a small fraction of the antenna's beam directivity will be non-zero below the horizon and pickup the temperature of the ground, which we define as ground loss. We simulated the antenna and ground plane using FEKO with a Green's function to implement the $30\times30$~m$^2$ ground plane as a perfect conductor over soil, similar to the technique used in \cite{moz01}.  We also simulated the antenna on an idealized infinite ground plane with FEKO and with CST Microwave Studio. The CST simulation resulted in a similar amount of loss as the FEKO simulation, but with an opposite frequency dependency as the FEKO simulation. Thus the compromise solution was to use a simplified ground loss function set to a constant 0.5\% loss, independent of frequency. This resulted in $\beta$ becoming more positive by up to 0.002 for 0~h~\textless~LST~\textless~12~h and by no more than 0.001 for 12~h~\textless~LST~\textless~24~h as compared to no ground loss correction.

\subsubsection{Antenna loss}
These losses develop along the balun, at the connection with the receiver, and from resistance in the antenna panels. The balun of the EDGES low-band system consists of two brass tubes (outer diameter 1.0 inch) with a copper-plated brass rod running up one of the tubes and a teflon-dielectric connector to the receiver. When we account for the balun and connector loss in the calibration process (estimated from transmission line models and lab measurements), we see an effect on $\beta$ no greater than 0.005. Hardware design changes in the low-band system, along with the lower frequency range, reduced the balun losses as compared to the high-band system. The effect on $\beta$ from the panel resistance is very minor at 0.001.

\subsubsection{Beam chromaticity}
\label{sec:BeamFactor}
\textcolor{black}{To compute the beam correction factor we used the beam created from the FEKO simulation with the realistic soil model and finite ground plane. We compared this beam against the beam resulting from FEKO simulations using an ideal infinite ground plane. Using one beam versus the other causes a difference in the spectral index of less than $\pm$0.004 for all LST values except from 20 to 23~h where it rises to +0.007. }

\textcolor{black}{We also estimate the effect of uncertainty in the spatial structure of the foreground sky at 75~MHz. Our standard method is to scale the Haslam 408~MHz sky map to 75~MHz using a single spectral index for the entire sky, after subtracting the CMB temperature. The spectral index value we recover from the corrected measurements is not sensitive to the choice of spectral index used in the scaling when it is in the range of -2.65 to -2.45. In order to further test the dependence of the recovered spectral index on the assumptions of the sky model, we considered three other scaling methods for projecting the Haslam sky map from 408 MHz to 75 MHz.  The methods were: 1)  Scale to 75 MHz using an index of -2.55 for Galactic latitudes within 10 degrees of the Galactic Plane (b=0) and -2.45 for all others; 2) Use the same method as in case 1, but within 20 degrees of the Galactic Plane; and 3) Scale to 75 MHz using the index from the \cite{guz01} sky map.  We find that the recovered spectral index varies by up to 0.01 relative to our nominal result when using these methods.
}

\subsubsection{Total uncertainty}
We estimate the total systematic uncertainty by assuming our adjustments have a standard error of half of the magnitude of the effect (with vs. without or method~1 vs method~2), and combine the sources of standard errors in quadrature as a function of LST: finite ground plane effects 0.006; antenna balun 0.005; finite panel resistance 0.001; \textcolor{black}{beam correction factor: a) directivity 0.007 and b) spatial structure 0.01, which are all maximum values but do not all occur at the same LST. This results in a systematic uncertainty on $\beta$ of~0.006}. The day to day scatter in beta per LST bin remains under 0.004 except for the LST values between 7 and 12~h, where it rises to 0.006 partly due to the difference in results from two instruments. Adding the data scatter standard deviation directly to the systematic uncertainty, as opposed to in quadrature, the total \textcolor{black}{maximum uncertainty in $\beta$ peaks at 0.011 for $8<$~LST~$<12$~h, 0.010 for $16<$~LST~$<20$~h} near the Galactic Centre, and $<$~0.008 for other LST values.  (see top graph in Fig.~\ref{fig:Grouped_Averages_two_params}).

When using a three-parameter fit, the systematic uncertainty in $\beta$ increases slightly for the ground loss and beam correction values. The ground loss uncertainty increases by 0.0005 to 0.0007 for LST values below 10 h and by 0.0002 to 0.0005  for LST values between 10 and 24 h. The systematic uncertainty in beam chromaticity is mostly the same except for LST values between 9.5 and 15~h where it becomes higher by up to 0.005. The maximum day to day scatter in beta per LST bin increases from 0.006 to 0.010 near LST~=~10~h. Combined, the total uncertainty in the spectral index for three-parameter fitting \textcolor{black}{increases to a maximum of 0.016 for $8<$~LST~$<12$~h, 0.011 for $17<$~LST~$<20$~h }near the Galactic Centre, and $<$~0.009 for other LST values.

\subsection{Ionosphere impact on the spectral index}
\label{sec:Ionosphere}
\begin{figure}
 \centering
 \includegraphics[trim=0 25 0 0, clip,width=3.3 in]{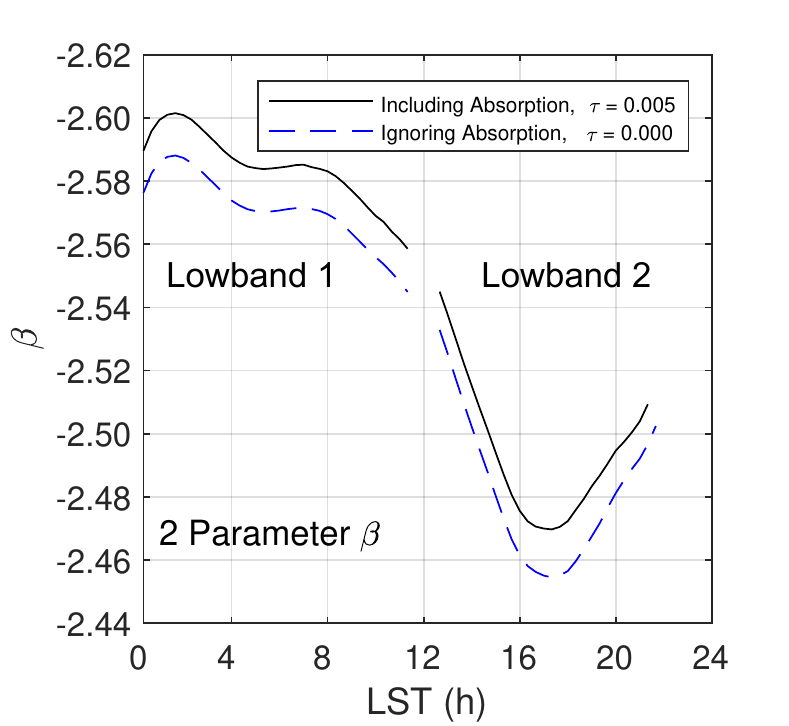}
 \includegraphics[trim=0 0 0 0, clip,width=3.3 in]{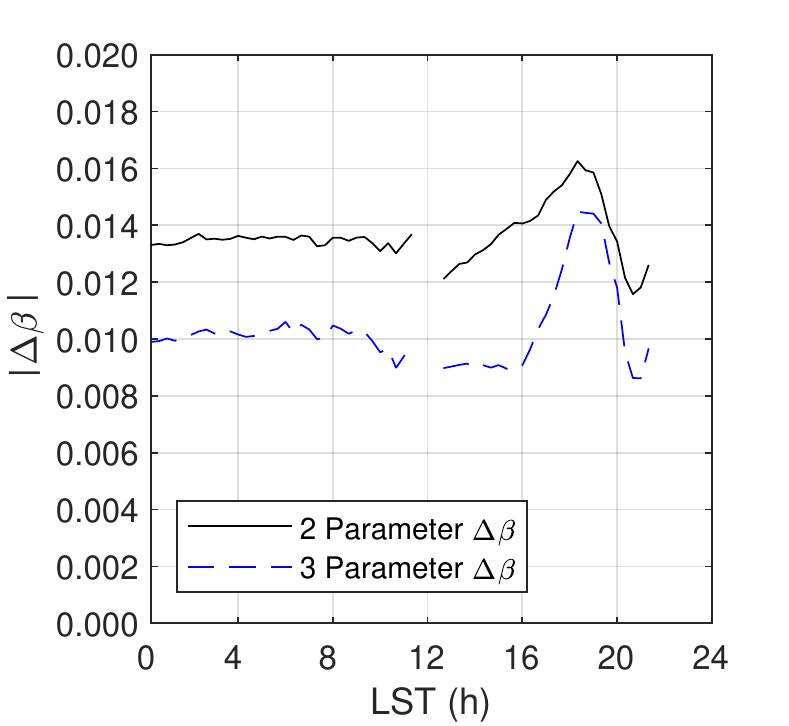}
 \caption{The top figure shows the effect of including a mild ionospheric absorption term (with $\tau=0.005$) in the two-parameter fitting equation for the spectral index, which causes the index to become more negative by an amount on the order of 0.012~to~0.016. The bottom figure shows the absolute difference between including and not including ionospheric absorption for both two and three-parameter fitting equations. The two data sets used were from Lowband 1 and Lowband 2 NS.}
 \label{fig:IonosphereDeltaBeta2}
\end{figure}

The ionosphere is expected to have an effect on the sky-averaged spectrum below 100 MHz \citep{ved01,sok01,dat01}.  When considering ionospheric absorption and emission, the extended two-parameter fitting equation is
\begin{multline} 
\label{eq:Power_Law_ion_full}
T_\text{sky} = T_{75}\left(\frac{\nu}{\nu_{_{75}}}\right)^{\beta}\times\left[e^{-\tau\left(\frac{\nu}{\nu_{_{75}}}\right)^{-2}}\right] \\+ T_{e}\left[1-e^{-\tau\left(\frac{\nu}{\nu_{_{75}}}\right)^{-2}}\right] + T_\text{CMB},
\end{multline} 
where $\tau$ is the coefficient of ionospheric absorption, and $T_e$ is the electron temperature in the ionosphere.  If we assume typical night-time values of $\tau\approx0.005$ and $T_e\approx1000$~K \citep{rog03}, the emission term (the second term in equation~(\ref{eq:Power_Law_ion_full})) becomes very minor, with less than 1~K contribution to the sky temperature across the observed band.  Thus, we can simplify the fitting equation by dropping the emission term and keeping only the first-order expansion of the absorption term, yielding
\begin{multline} 
\label{eq:Power_Law_ion}
T_\text{sky} = T_{75}\left(\frac{\nu}{\nu_{_{75}}}\right)^{\beta}\times\left[1-\tau\left(\frac{\nu}{\nu_{_{75}}}\right)^{-2}\right]  + T_\text{CMB}.
\end{multline} 
A similar equation can be constructed for the three-parameter power-law model by including the $\gamma$ factor as in equation~(\ref{eq:Power_Law_3P}).  Fitting equation~(\ref{eq:Power_Law_ion}) and the three-parameter counterpart to the data using a fixed $\tau=0.005$, we find that including ionospheric absorption in the fit makes the spectral index more negative (steeper) by a relatively constant amount across LST for both two and three-parameter fitting. The differences are shown in Fig.~\ref{fig:IonosphereDeltaBeta2} and range from 0.008 - 0.016, with the larger offsets occurring near the Galactic Centre where the power law profile is shallowest and the spectral index is greatest. For $\tau=0.005$, the ionospheric absorption term ranges from 0.22\% at 50~MHz to 0.89\% at 100~MHz. At 75~MHz, the effect is 0.5\% and the minimum and maximum values of $T_{75}$ across LST are decreased by 8~and~23~K, respectively.   We verified that omitting the emission term was justified by fitting equation~(\ref{eq:Power_Law_ion_full}) using a fixed value of $T_e=1000$ and confirming that no significant changes arose to the spectral index values compared to fits with the simplified form.

\subsection{Extended model}
\label{sec:extended}
To investigate the possibility of bias in our parameter estimates due to insufficient degrees of freedom in the model, we extended the three-parameter model of equation~(\ref{eq:Power_Law_3P}) to five free parameters. The new model is given by
\begin{equation} 
\label{eq:Power_Law_5P}
T_\text{ant} = T_{75}\left(\frac{\nu}{\nu_{_{75}}}\right)^{\beta+\gamma \ln(\frac{\nu}{\nu_{_{75}}})+a_4[\ln(\frac{\nu}{\nu_{_{75}}})]^2+a_5[\ln(\frac{\nu}{\nu_{_{75}}})]^3} + T_\text{CMB},
\end{equation} 
where the new parameters $a_4$ and $a_5$ accompany the higher orders of $\ln(\frac{\nu}{\nu_{_{75}}})$ in the exponent of $\left(\frac{\nu}{\nu_{_{75}}}\right)$. We find that these additional parameters have very little effect on the fitted parameters to the three-parameter equation other than lowering the RMS residual as can be expected from an increase in available degrees of freedom.  These results are summarized in Table \ref{tab:Fitting_Parameter_Summary}.

\subsubsection{Residuals for two- and three-parameter fitting}
\begin{figure}
\centering
   \includegraphics[trim=0 20 0 0, clip,width=3.3 in]{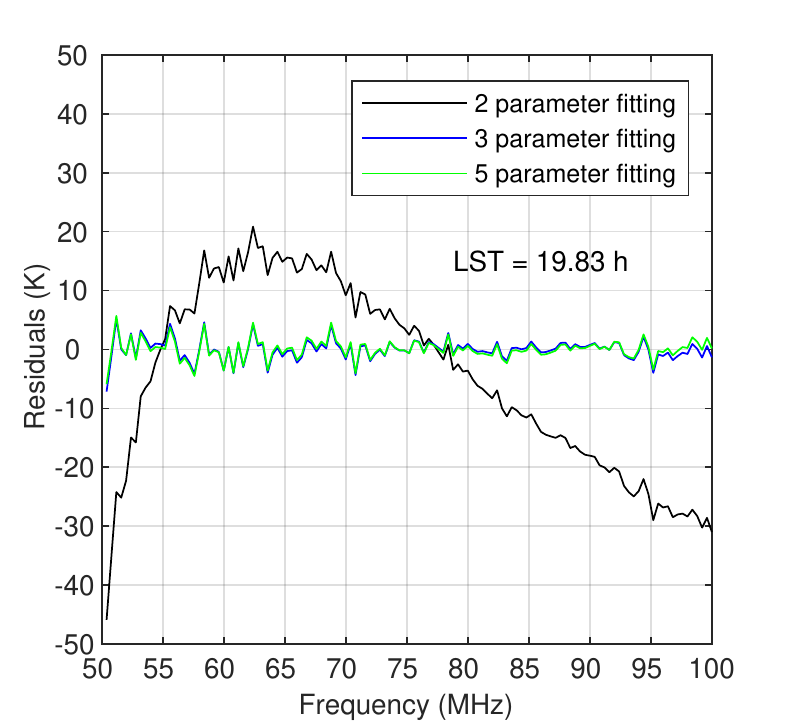}
   \includegraphics[trim=0   0 0 0, clip,width=3.3 in]{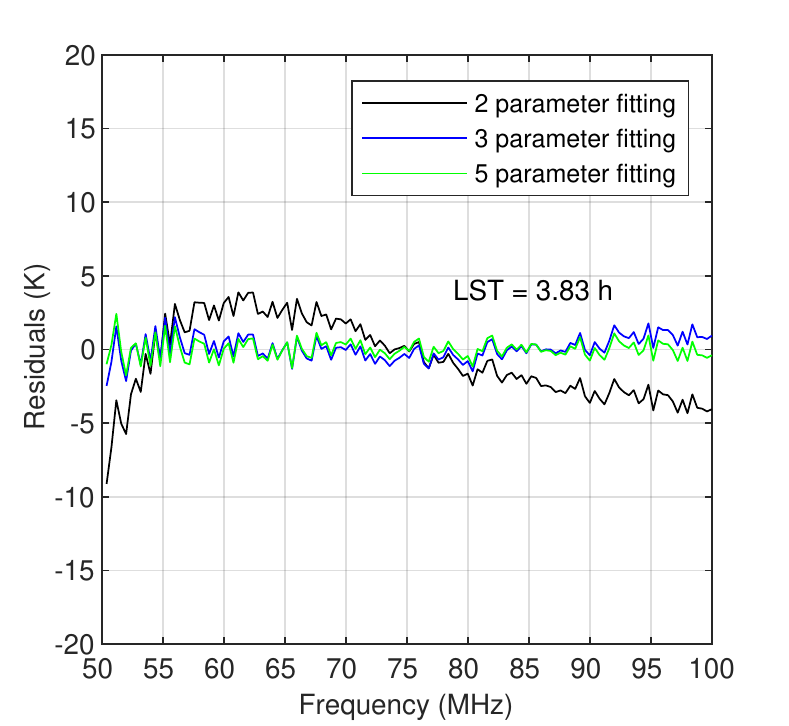}
   \caption{\textcolor{black}{Residuals to two, three, and five parameter fittings of two representative 20-minute observations for high-residual (top) and low-residual (bottom) regions at LST=19.83~h and 3.83~h, respectively. The data are from the Lowband 1 NS instrument on day 264 and have spectral channels of 400~kHz. The residual RMS values for two, three, and five parameter fitting at LST=19.83~h are 16, 1.8, and 1.7~K, and at LST=3.83 h are 2.7, 0.85, and 0.66~K, respectively. The residuals in the two-parameter fittings are not dominated by thermal noise, but instead by smooth spectral structure. In contrast, the residuals of the three-parameter fittings are much closer to the thermal noise limit.  The five-term model discussed in section \ref{sec:extended} yields completely noise-like residuals at the sensitivity of the 20 minute observations.}}

\label{fig:Residuals}
\end{figure}

\textcolor{black}{The residuals of the fittings for the spectral index were examined for two, three, and five parameter fittings for both low and high residual locations in LST. In Fig. \ref{fig:Residuals} we show, for reference, the 400~kHz residuals from day 264. The residual RMS values for two, three, and five parameter fitting at LST=19.83~h are 16, 1.8, and 1.7~K, and at LST=3.83~h are 2.7, 0.85, and 0.66~K, respectively. The residuals in the two-parameter fittings are not dominated by thermal noise, but instead by smooth spectral structure. In contrast, the residuals of the three-parameter fittings are much closer to the thermal noise limit. The five-term model yields completely noise-like residuals at the sensitivity of the 20 minute observations.}

\subsection{Summary of fitting results}
\label{sec:summary}
The first two columns of Table \ref{tab:Fitting_Parameter_Summary} summarize the two and three-parameter fitting result averages of the 244 nights of collected data while ignoring ionospheric effects. The next two columns show the estimated effect on the averages when including the expected typical amount of night-time ionospheric absorption, and the last column lists the results for five-term fitting. The general trend is that the two-parameter model results in a more negative spectral index than the three-parameter model ranging from 0.014 to 0.029 depending upon the specific LST value, and the difference between three and five parameter fitting is less than 0.003. Accounting for a small amount of ionospheric absorption also tends to make the spectral index more negative by 0.008 - 0.016. The RMS residual improves as we increase the number of fitting parameters from two to three, but appears to be near saturation as the difference between the five parameter RMS residual and the three-parameter RMS residual is $< 25$\% compared to the decrease by factors of 3 to 5 for the two-parameter to three-parameter RMS residual reduction. $T_{75}$ is nearly constant for a given LST, varying by less than 0.6\% across the various fitting styles. The spectral index curvature, $\gamma$, ranges between $-0.04$ and $-0.11$ (without ionospheric absorption) which is more positive than, but not inconsistent with, values reported by \cite{deo01}.
\begin{table}
   \caption{Fitting parameter values when averaged over all days of all data sets at four LST values. For two and three-parameter fitting, we show the values for the case of no ionospheric absorption and mild ($\tau=0.005$) ionospheric night-time absorption.  We also show results for fitting to a five term exponential-log equation without ionospheric absorption. Antenna temperature values were placed into 72 LST bins (each representing a \textcolor{black}{20 minute observation}) and 400~kHz frequency bins before fitting.}
   \label{tab:Fitting_Parameter_Summary}
   \begin{tabular} {l c c c c c c }
   	\hline
Para-&LST& \multicolumn{2}{c}{no ionospheric}&\multicolumn{2}{c}{with ionospheric}& Exp-log\\
meter&(h)& \multicolumn{2}{c}{corrections}&\multicolumn{2}{c}{corrections}& \\
&& \multicolumn{2}{c}{(fitting terms)}&\multicolumn{2}{c}{(fitting terms)}&(terms)\\
&& 2 &3 & 2& 3&5\\
\hline
$T_{75}$   & 0 & 1806 & 1807 & 1815 & 1816 & 1807\\
(K)            & 6 & 1673 & 1673 & 1681 & 1682 & 1673\\
               & 12& 2566 & 2568 & 2579 & 2580 & 2568\\
               & 18& 4749 & 4752 & 4773 & 4776 & 4751\\
\hline
$\beta$   & 0 & -2.576 & -2.592 & -2.590 & -2.603 &-2.591\\
             & 6 & -2.571 & -2.585 & -2.585 & -2.595 &-2.585\\
             & 12& -2.539 & -2.568 & -2.553 & -2.578 &-2.565\\
             & 18& -2.463 & -2.489 & -2.477 & -2.499 &-2.489\\
\hline
$\gamma$   & 0 & - & -0.055 & - & -0.042  & -0.068\\
                 & 6 & - & -0.047 & - & -0.034  & -0.041\\
                 &12 & - & -0.099 & - & -0.086 & -0.090\\
                 &18 & - & -0.089 & - & -0.076 &-0.079\\
\hline
$a_4$         & 0 & - & - & - & -  & -0.048\\
                 & 6 & - & - & - & -  & -0.004\\
                 &12 & - & - & - & - & -0.053\\
                 &18 & - & - & - & - & +0.018\\
\hline
$a_5$         & 0 & - & - & - & -  & -0.022\\
                 & 6 & - & - & - & -  & -0.031\\
                 &12 & - & - & - & - & -0.158\\
                 &18 & - & - & - & - & -0.025\\
\hline
RMS      & 0 &  3.7 & 1.2 & 2.9 & 1.2 &1.0\\
resid.    & 6 &  2.9 & 0.9 & 2.2 & 0.9 &0.9\\
(K)      & 12&  9.0 & 1.6 & 7.9 & 1.6 &1.4\\
          & 18& 15   & 3.6 & 13 & 3.6 &2.8\\
   	\hline
		\end{tabular}
\end{table}

\subsection{Simulated spectral index from sky maps} 
\label{sec:sim}
\begin{figure}
 \centering
 \includegraphics{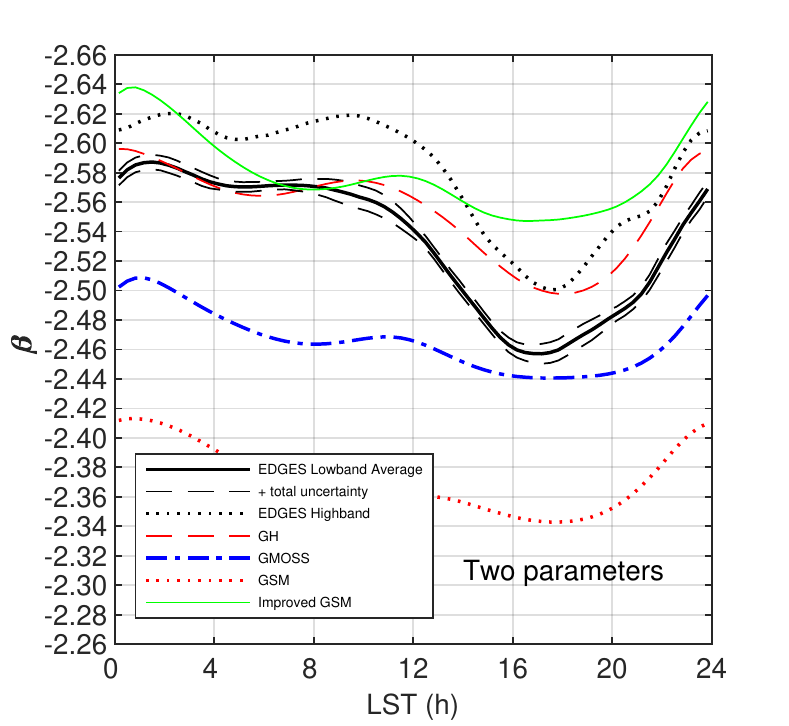}
 \includegraphics{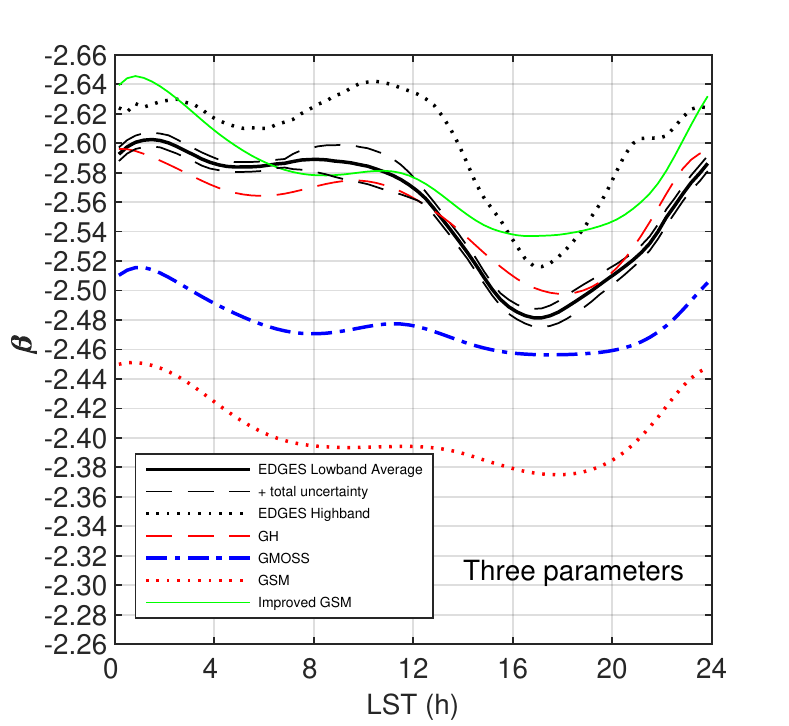}
 \caption{Two-parameter (top) and three-parameter (bottom) derived spectral index measurements from EDGES low-band and high-band instruments and from simulated spectral index results using the GH-model (Guzm\'an and Haslam maps at 45 MHz and 408 MHz, respectively), the GMOSS, the GSM, and the improved GSM skymaps. The GH-model shows the best agreement with the three-parameter low-band measured data results, differing by at most 0.02. The GH-model agrees with the two-parameter low-band measured results better at low LST values and agrees better with the three-parameter high-band measured results better at higher LST values. The improved GSM agrees better with measured results than the original GSM and the GMOSS model, and agrees better with three-parameter measured results than the two-parameter results. The uncertainty band includes data scatter and systematic errors.}
 \label{fig:GSM_vs_Guz-Has}
\end{figure}

We compare our measured spectral index results (without including the ionospheric absorption in the models) to values obtained by simulating observations with the EDGES beam (NS orientation) and several publicly available sky maps: the de Oliveira-Costa GSM \citep{deo01}; the improved GSM \citep {zhe01}; the GMOSS \citep{rao01}; and the Haslam 408~MHz map \citep{has01,has02} in conjunction with the Guzm\'an 45~MHz map \citep{guz01}. We simulate the antenna temperature that EDGES is expected to observe using multifrequency sky maps from 50 to 100~MHz in steps of 1~MHz as a function of LST, by convolving the sky maps, $T^{'}_\text{sky-model}(\nu,\Omega)$ (less $T_{\text{CMB}}$) with the simulated EDGES low-band antenna beam (fixed at 75~MHz to neglect beam chromaticity effects since we correct for beam chromaticity in our primary analysis).   For the simulated observations, we use
\begin{equation} 
\label{eq:T_antenna}
T^{'}_\text{ant}(\nu) = \int_{\Omega}T^{'}_\text{sky-model}(\nu,\Omega) B(\nu_{_{75}},\Omega) \mathrm{d}\Omega + T_{\text{CMB}},
\end{equation}
where $B(\nu_{_{75}}, \Omega)$ is the beam directivity at 75~MHz for a given pointing and orientation, $\nu$ is frequency, $\Omega$ are the spatial coordinates above the horizon,  and $T^{'}_\text{sky-model}(\nu,\Omega)$ is the sky model in use. $B(\nu_{_{75}}, \Omega)$ is normalized to a unit integral. 
The simulated antenna temperature is then fit to the two-parameter equation~(\ref{eq:Power_Law_2P}) and the three-parameter equation~(\ref{eq:Power_Law_3P}) to obtain the spectral index vs. LST.  Ionospheric effects are ignored in all cases. We use the Guzm\'an and Haslam sky maps  at 45~MHz and 408~MHz as a pair (GH) to arrive at a self-consistent spectral index map using the closed form expression
\begin{equation} 
	\label{eq:Two_frequency_spectral_index}
	\beta =\text{ln}\left(\frac{T^{'}_{\text{ant}}(\nu_{_{45}})}{T^{'}_{\text{ant}}(\nu_{_{408}})}\right) \bigg/ \text{ln}\left(\frac{\nu_{_{45}}}{\nu_{_{408}}}\right),
\end{equation} 
In this manner, we generate spectral index values across 24 hours of LST. The results of these simulations, along with the low-band and high-band measurements, are displayed in Fig.~\ref{fig:GSM_vs_Guz-Has}.  

For two-parameter fitting, the GH-model shows good agreement with measurements at low LST values, but the spectral index becomes more negative than the measurements by up to 0.04 around the Galactic Centre. For three-parameter fitting, the GH-model, which has no intrinsic spectral index curvature ($\gamma=0$), shows more consistent agreement with measurements of spectral index across all LST values, differing by only up to $\pm0.02$ across all LST. The GMOSS and improved GSM models bracket the measured data, with the improved GSM model more negative than the measured values, while the GMOSS model yields more positive predictions of the spectral index. We also include the spectral index as reported in the high-band paper \citep{moz01}. In general, the measured low-band spectral index has become less negative (flattened) by approximately 0.03 to 0.06 as compared to the high-band results. Ionospheric absorption effects are not included in the graphs, but the analysis in section~\ref{sec:Ionosphere} indicates that ionospheric absorption likely only accounts for a change of $\lesssim0.02$ in $\beta$, and assuming that the ionospheric effects are lower in the high-band data, the intrinsic spectral index flattens at most by ($\lesssim0.02$) from 90-190~MHz to 50-100~MHz. 

\section{CONCLUSION}
\label{sec:conclusion}
We measured the sky brightness temperature as a function of frequency ($50-100$~MHz) using two implementations of the EDGES low-band instrument and derived the spectral index $\beta$ as a function of sidereal time by fitting to two-parameter and three-parameter equations using 244 days of night-time data acquired from 14 September 2016 to 27 August 2017. Instrument calibration, including corrections for ground loss, antenna losses, and beam chromaticity have been applied to deliver instrument stability over several months as demonstrated by spectral index standard deviation values of $\sigma_{\beta}~<~0.006$ for two-parameter fitting and $\sigma_{\beta}~<~0.010$ for three-parameter fitting.

For two-parameter fitting, the spectral index $\beta$ is in the range $-2.59$~\textless~$\beta$~\textless~$-2.54~\pm$~\textcolor{black}{0.011} in the lower LST values corresponding to high Galactic latitudes, but increases to $-2.46$ near LST=17.6 h where the Galaxy is overhead. The measured low-band spectral index is less negative by approximately 0.03 to 0.06 as compared to the previously reported high-band ($90-190$~MHz) results. We find that the GMOSS-derived spectral indices are more positive than the measured results especially for low LST values (up to $+0.10$). The improved GSM model results are more negative than the measured results especially near the Galactic Centre (up to $-0.10$).  The Guzm\'an-Haslam derived spectral indices, on the other hand, match two-parameter measurements to within 0.01 at low LST values and within 0.04 near the Galactic Centre.  For three-parameter measurements, the GH values are within $\pm0.02$ across all LST values. Three-parameter spectral index values are more negative than two-parameter measured values by approximately 0.02 and the total maximum \textcolor{black}{uncertainty increases to 0.016}.  We also find that the third parameter, the spectral index curvature, $\gamma$, is constrained to $-0.11<\gamma<-0.04$. Including night-time ionospheric absorption, using $\tau=0.005$, makes the spectral index more negative across LST by an amount ranging from 0.008 - 0.016.

\section*{ACKNOWLEDGMENTS}
This work was supported by the NSF through research awards for the Experiment to Detect the Global EoR Signature (AST-1207761 and AST-1609450). R.A.M. was supported by the NASA Solar System Exploration Virtual Institute cooperative agreement 80ARC017M0006 and by the NASA Ames Research Center grant NNX16AF59G. EDGES is located at the Murchison Radio-astronomy Observatory. We acknowledge the Wajarri Yamatji people as the traditional owners of the Observatory site. We thank CSIRO for providing site infrastructure and support.

\bsp

\label{lastpage}
\end{document}